\documentclass[11pt]{article}
\usepackage{fancyhdr}
\usepackage{isomath}
\usepackage{amsmath}
\usepackage{amsbsy}
\usepackage{amssymb}
\usepackage{amscd}
\usepackage{amsfonts}
\usepackage{graphicx,color}
\usepackage{verbatim}
\usepackage{euscript}
\usepackage{alltt}
\usepackage{stmaryrd}
\usepackage{subfigure}

\newenvironment{rcases}
  {\left.\begin{aligned}}
  {\end{aligned}\right\rbrace}

\usepackage{amsmath}
\usepackage{amsbsy}
\usepackage{amssymb}
\usepackage{amscd}
\usepackage{amsfonts}

\newcommand{\R}{\mathbb R}

\newcommand{\bfn}{{\mathbold n}}

\newcommand{\bfx}{{\mathbold x}}

\newcommand{\bfC}{{\mathbold C}}

\newcommand{\bfF}{{\mathbold F}}

\newcommand{\bfR}{{\mathbold R}}

\newcommand{\bfT}{{\mathbold T}}
\newcommand{\bfU}{{\mathbold U}}

\newcommand{\bfW}{{\mathbold W}}

\newcommand{\beq}{\begin{equation}}
\newcommand{\eeq}{\end{equation}}
\newcommand{\beqs}{\begin{eqnarray}}
\newcommand{\eeqs}{\end{eqnarray}}
\newcommand{\beql}{\begin{equation} \label}
\newcommand{\half}{\frac{1}{2}}

\newcommand{\bfalpha}{\mathbold{\alpha}}
\newcommand{\bfomega}{\mathbold{\omega}}

\newcommand{\bfveps}{\mathbold{\varepsilon}}
\newcommand{\bfeta}{\mathbold{\eta}}

\newcommand{\bfzero}{\mathbf{0}}

\newcommand{\grad}{\mathop{\rm grad}\nolimits}

\newcommand{\curl}{\mathop{\rm curl}\nolimits}

\newcommand{\parderiv}[2]{\frac{\partial #1}{\partial #2}}

\usepackage[margin=1in]{geometry}

\date{September 6, 2018}
\begin{document}
\title{Stress of a spatially uniform dislocation density field}
\author{Amit Acharya\thanks{Dept. of Civil \& Environmental Engineering, and Center for Nonlinear Analysis, Carnegie Mellon University, Pittsburgh, PA 15213, email: acharyaamit@cmu.edu.} 
}
\maketitle

\begin{abstract}
\noindent It can be shown that the stress produced by a spatially uniform dislocation density field in a body comprising a linear elastic material under no loads vanishes. We prove that the same result does not hold in general in the geometrically nonlinear case. This problem of mechanics establishes the purely geometrical result that the $\curl$ of a sufficiently smooth two-dimensional rotation field cannot be a non-vanishing constant on a domain.
\end{abstract}

\section{Introduction}
Dislocations are defects of compatibility in an elastic solid that often produce stress in the absence of applied loads. However, there are many dislocation distributions that produce no stress, at least in the limit of a linear or nonlinear continuum elastic description \cite{mura1989impotent, head1993equilibrium, yavari2012riemann}. Such distributions are physically important in the far-field description of grain boundaries, in understanding the resistance produced to dislocation motion resulting in plastic deformation, and for questions of patterning and microstructure in plastic deformation. To be specific, all dislocation distributions resulting from a curl of any (linearized) rotation field on a domain form the collection of stress-free dislocation distributions on that domain (for an appropriate class of stress response functions) - clearly this is a large class of fields, both in the linear and nonlinear settings. An interesting result in the linear theory of dislocations, that we give a proof of in Sec. \ref{sec:lin}, is that any spatially uniform dislocation density field belongs to this class of stress-free dislocation density fields. To our knowledge, whether such a result holds in the nonlinear elastic theory of dislocations is not known. We prove such a result in the negative in Sec. \ref{sec:nlin} in the setting of two space dimensions. 

The paper is organized as follows: in Sec. \ref{sec:lin} we provide an independent proof of the result in the linear case. Sec. \ref{sec:nlin} discusses the nonlinear case. We use standard tensor notation. All tensor and vector components are written with respect to the basis of a fixed Rectangular Cartesian coordinate system that is assumed to parametrize  ambient 3-d space with a generic point denoted as $(x_1, x_2, x_3)$. A subscript comma represents partial differentiation and the summation convention is always used. The curl of a tensor field is represented in components by row-wise curls of the corresponding matrix of the tensor field w.r.t the basis of the Rectangular Cartesian coordinate system in use.

The resolution of the question in space dimension 3 for constant dislocation density fields without any further  restriction is left for future work, as is exploring the possibility of necessary conditions, solely in terms of the dislocation density and its higher derivatives, for the dislocation density to be a curl of a finite rotation field. Of interest is also the question of characterizing stress-free dislocation density distributions on a domain in situations when the elastic stress response function of the material is non-monotone, and is known to admit a set of elastic distortions at which the stress response function evaluates to zero.

\section{The result in the linear case}\label{sec:lin}
We assume $\Omega$ to be a simply connected domain on which is specified a dislocation density field $\bfalpha$. The body comprises a generally non-homogeneous, linear elastic material, whose elastic modulus, $\bfC$, has major and minor symmetries. The governing equations for the `internal' stress field in the body are given by
\begin{equation}\label{eqn:gov_Ue}
\begin{split}
\begin{rcases}
&\curl \bfU  = \bfalpha \\
&\mbox{div} \left( \bfC \bfU \right) = \bfzero
\end{rcases}
\qquad \bfx\in\Omega,\\
\left( \bfC \bfU \right) \bfn = \bfzero \qquad \bfx \in \partial \Omega,
\end{split}
\end{equation}
where $\bfU$ is the elastic distortion and $\bfn$ is the outward unit normal field on the boundary of $\Omega$.

The symmetric part, $\bfveps := \half \left(\bfU + \bfU^T \right)$, of any solution of \eqref{eqn:gov_Ue}$_1$ satisfies the relation
\begin{equation}\label{eqn:strn_inc}
\left( \curl \left( (\curl \bfU)^T \right) \right)_{sym} = \curl \left( (\curl \bfveps)^T \right) = (\curl \bfalpha^T)_{sym} =: \bfeta,
\end{equation}
where $\bfeta$ is Kr\"{o}ner's incompatibility tensor. Equation \eqref{eqn:strn_inc} implies that the strain field of any solution of \eqref{eqn:gov_Ue} corresponding to a prescribed dislocation density with a vanishing incompatibility field is strain-compatible, i.e., it is the symmetrized gradient of a vector field on $\Omega$, \eqref{eqn:strn_inc} then simply becoming the St.-Venant compatibility condition for $\bfveps$. But then \eqref{eqn:gov_Ue}$_{2,3}$ imply that $\bfveps = \bfzero$ (by Kirchhoff's uniqueness theorem for linear elastostatics), and consequently the stress of such a dislocation density field vanishes. Of course, $\bfU$ in such cases is a skew tensor field, and spatially non-uniform for $\bfalpha \neq \bfzero$. For $\bfalpha = \bfzero$, $\bfU$ is necessarily a constant skew tensor field on $\Omega$. The first equality in \eqref{eqn:strn_inc} suggests that $\left( \curl \left( (\curl \bfomega)^T \right) \right)_{sym} = \bfzero$ for any skew-symmetric tensor field $\bfomega$, so that a necessary condition for a dislocation density field to be a curl of a skew-symmetric tensor field is that its incompatibility $(\bfeta)$ field vanishes. These results hold for the general three-dimensional situation as well as for two-dimensional plane problems.

Clearly, a spatially uniform dislocation density field has vanishing incompatibility, and therefore has a stress field that vanishes.

\section{The nonlinear case}\label{sec:nlin}
We consider a generally nonlinear stress response function $\bfT(\bfF)$ taking invertible tensors as arguments and with the property that $\bfT = \bfzero$ if and only if $\bfF$ is an orthogonal tensor. We shall denote $\bfF^{-1} =: \bfW$. The governing equations for 
the `internal' stress field in the body in the nonlinear case are given by \cite{willis1967second}
\begin{equation}\label{eqn:gov_nonl}
\begin{split}
\begin{rcases}
&\curl \bfW  = - \bfalpha \\
&\mbox{div} \left( \bfT(\bfF) \right) = \bfzero
\end{rcases}
\qquad \bfx\in\Omega,\\
\left( \bfT(\bfF) \right) \bfn = \bfzero \qquad \bfx \in \partial \Omega,
\end{split}
\end{equation}
where $\bfF$ is the elastic distortion and, as before, $\bfalpha$ is considered as prescribed.

Henceforth, we restrict attention to the class of specified dislocation density tensor fields that have $\alpha_{13}$ and $\alpha_{23}$ as the only possible non-vanishing components that are constant on $\Omega$. Hence, such a field is spatially uniform and we wish to determine whether there exists a $C^2$, orthogonal tensor-valued solution to \eqref{eqn:gov_nonl}$_1$, which would correspond to a stress-free solution to \eqref{eqn:gov_nonl}$_{2,3}$. We will also restrict attention to fields that do not vary in the $x_3$ (out-of-plane) direction in $\Omega$.

\emph{Restricting attention to planar $C^2(\Omega)$ solutions of \eqref{eqn:gov_nonl}$_1$ that correspond to rotations in the $x_1-x_2$ plane}, we show in the following that \emph{there does not exist $\theta \in C^2(\Omega, \R)$ such that $\curl \bfR (\theta) = -\bfalpha$ for any constant field $\bfalpha \neq \bfzero$} through a proof by contradiction, where the components of $\bfR (\theta)$ are defined in \eqref{eqn:rot_comp} below.

Let us assume that a $\theta \in C^2(\Omega, \R)$ exists corresponding to a planar rotation-valued solution, $\bfR$, to \eqref{eqn:gov_nonl}$_1$ for some constant $\bfalpha \neq \bfzero$ field on $\Omega$. The matrix representation of  $\bfR$ with respect to the fixed orthonormal basis corresponding to the $x_1-x_2-x_3$ directions is given by
\begin{equation}\label{eqn:rot_comp}
[ R(\theta) ] = \begin{bmatrix}
\cos \theta & - \sin \theta & 0\\
\sin \theta & \cos \theta & 0\\
0 & 0 & 1
\end{bmatrix}.
\end{equation}
The matrix $R$ satisfies
\begin{equation*}
e_{3jk} R_{mk,j} = - \alpha_{m3} \Longrightarrow e_{312} R_{m2,1} + e_{321} R_{m1,2} = - \alpha_{m3} \qquad m = 1,2,
\end{equation*}
where $e_{ijk}$ is a component of the alternating tensor, which further implies that
\begin{equation*}
\begin{bmatrix}
-\cos \theta & \sin \theta\\
-\sin \theta & - \cos \theta
\end{bmatrix}
\begin{bmatrix}
\theta_{,1}\\
\theta_{,2}
\end{bmatrix}
=
- \begin{bmatrix}
\alpha_{13}\\
\alpha_{23}
\end{bmatrix}
\end{equation*}
so that
\begin{equation}\label{eqn:gradtheta}
\theta_{,i} = A_{ij}(\theta) a_j, \qquad i, j = 1, 2
\end{equation}
where 
\begin{equation*}
A(\theta) =
\begin{bmatrix}
-\cos \theta & - \sin \theta\\
\sin \theta & - \cos \theta
\end{bmatrix}; \qquad 
a = \begin{bmatrix}
-\alpha_{13}\\
-\alpha_{23}
\end{bmatrix}.
\end{equation*}
Since $\theta$ is $C^2(\Omega)$, we have $\theta_{,im} = \theta_{,mi}$ which implies
\begin{equation*}
\parderiv{A_{ij}}{\theta} \theta_{,m} a_j + A_{ij} a_{j,m} - \parderiv{A_{mj}}{\theta} \theta_{,i} a_j - A_{mj} a_{j,i} = 0  \Longrightarrow \parderiv{A_{ij}}{\theta} A_{mp} a_p a_j - \parderiv{A_{mj}}{\theta} A_{ip} a_p a_j = 0,
\end{equation*}
since $\bfalpha$ is spatially uniform by hypothesis. Thus we have
\begin{equation}\label{eqn:symm}
\left( \parderiv{A}{\theta} a \right) \otimes (A a) - (A a) \otimes \left( \parderiv{A}{\theta} a \right) = 0 \quad \Longrightarrow \quad \left( \parderiv{A}{\theta} a \right) \otimes (A a) \  \mbox{ is symmetric},
\end{equation}
where
\begin{equation*}
\parderiv{A}{\theta} a =
\begin{bmatrix}
a_1 \sin \theta - a_2 \cos \theta\\
a_1 \cos \theta + a_2 \sin \theta
\end{bmatrix}; \qquad
Aa = 
\begin{bmatrix}
-a_1 \cos \theta -a_2 \sin \theta\\
a_1 \sin \theta - a_2 \cos \theta
\end{bmatrix}.
\end{equation*}
Consequently, we must have
\begin{equation*}
(a_1 \sin \theta - a_2 \cos \theta)^2  =  -(a_1 \cos \theta + a_2 \sin \theta)^2
\end{equation*}
which implies
\begin{equation}\label{eqn:implic}
\begin{split}
a_1 \sin \theta - a_2 \cos \theta & = 0\\
a_1 \cos \theta + a_2 \sin \theta  & = 0.
\end{split}
\end{equation}
$a_1 = a_2 = 0$ is not allowable by hypothesis. If $a_1 \neq 0$, $\sin \theta = \frac{a_2}{a_1} \cos \theta$ and $\cos \theta \left[ 1 + \left(\frac{a_2}{a_1}\right)^2 \right] = 0$ which implies the absurdity that $\cos \theta = 0$ and $\sin \theta = 0$. The same absurd conclusion is reached if $a_2 \neq 0$ which implies  $\cos \theta = \frac{a_1}{a_2} \sin \theta$ and $\sin \theta \left[ 1 + \left(\frac{a_1}{a_2}\right)^2 \right] = 0$.
Consequently, it must be true that \emph{planar rotation solutions to \eqref{eqn:gov_nonl}$_1$ parametrized by a $\theta \in C^2(\Omega, \R)$ field cannot exist for any constant field $\bfalpha \neq \bfzero$} (with only $\alpha_{13}$ and $\alpha_{23}$ as possible non-zero components). 

Thus, a non-vanishing, \emph{constant} dislocation density field comprising straight edge dislocations with arbitrary Burgers vector in the plane normal to the line direction cannot be stress-free when the allowed class of elastic distortions are `planar' fields varying only in $(x_1, x_2)$ and satisfying $F_{31} = F_{32} = F_{13} = F_{23} = 0$ and $F_{33} = 1$. Of course, it is easy to construct non-constant, dislocation density fields comprising straight edge dislocations that are stress-free, both in the linear and nonlinear settings.

Our result also directly proves that the $\curl$ of a rotation in two space dimensions cannot be a nonvanishing constant.

\section*{Acknowledgments}
It is a pleasure to acknowledge discussions with Reza Pakzad, who was able to construct a proof of the result presented here in a setting with weaker regularity. I acknowledge the support of the Center for Nonlinear Analysis at Carnegie Mellon and grants ARO W911NF-15-1-0239 and NSF-CMMI-1435624.
\bibliographystyle{alpha}\bibliography{nlin_stress_free}

\end{document}